1    Network structure and phylogenetic signal in an artificially assembled

2    plant-pollinator community

7    **James E. Cresswell[1*] and Laura Hernandez[2]**

11    [1]*Biosciences, College of Life & Environmental Sciences, University of Exeter,*

12    *Hatherly Laboratories, Prince of Wales Road, Exeter, EX4 4PS, United Kingdom;*

13    [2]*Laboratoire de Physique Théorique et Modélisation, UMR8089 CNRS-University of*

14    *Cergy-Pontoise, 2 Avenue Adolphe Chauvin, 95302, Cergy-Pontoise Cedex, France;*

16    ORIGINAL ARTICLE

18    Running title: Structure of an artificial community

20    [*]Author for correspondence (j.e.cresswell@ex.ac.uk)

1       ABSTRACT

3   • *Background and Aims*  Community ecologists are principally occupied with the

4       proposition that natural assemblages of species exhibit orderliness and with

5       identifying its causes.  Plant-pollinator networks exhibit a variety of orderly

6       properties, one of which is 'nestedness'.  Nestedness has been attributed to

7       various causes, but we propose a further influence arising from the phylogenetic

8       structure of the biochemical constraints on the pollen diets of bees.  We use an

9       artificial assemblage as an opportunity to isolate the action of this mechanism.

10  • *Methods*   We analyzed a network describing interactions between 52 bee

11      species and a collection of milkvetches (*Astragalus* spp.) and sainfoins

12      (*Onobrychis* spp.) in a botanical garden in the northwest of the United States of

13      America, whose papilionoid flowers were relatively similar in form.

14  • *Key Results*  The network connections of both plants and pollinators were

15      significantly nested.  The network both carried phylogenetic signal, whereby

16      related species tend to interact with the same mutualists, and exhibited

17      phylogenetic reciprocy, whereby the partners of evolutionarily related species

18      were, themselves, related.

19  • *Conclusions*  The properties of the network that we studied are consistent with

20      the proposition that nestedness is caused by the phylogeny of diet range in bees,

21      but the claim is preliminary and we propose that valuable progress in

22      understanding plant-pollinator systems may be made through applying the

23      techniques of chemical ecology at the community scale.

1    **Key words**: chemical ecology, community structure, oligolecty, pollen, polylecty,

2    social bees, solitary bees

1    INTRODUCTION

3    Community ecologists are principally occupied with the proposition that natural

4    assemblages of species exhibit orderliness (Hutchinson, 1953).  Formally, the

5    orderliness of a natural assemblage can be established by comparison with randomly

6    assembled communities, which are produced by computer algorithms (Connor and

7    Simberloff, 1979).  If interactions among species are depicted as web diagrams

8    (Elton, 1927; Lindeman 1942), orderliness can be sought in the patterns that become

9    apparent (Briand and Cohen).  For webs that depict trophic relationships, or food

10   webs, various generalizations emerge (Dunne, 2005).  For example, food chains are

11   typically short and cycles are rare (Cohen, 1989).  It is natural to ask after the causes

12   of these patterns.  Some recurrent structural motifs that are attributable to their

13   relative stability when interactions affect the population dynamics of species (May,

14   1972), but it is important also to recognize that interactions between species are

15   based on individual-level mechanisms.  For example, predator-prey relations depend

16   greatly on size relationships (whether it is feasible for a predator to subdue and eat a

17   certain prey) and on the economic decisions made by predators (whether it is

18   worthwhile for a predator to eat a certain prey; Elton, 1927) and some important

19   structural motifs of food webs are well explained by this (Beckerman *et al.*, 2006;

20   Petchey *et al.*, 2008).  Mechanisms that arise from the intrinsic morphological and

21   physiological attributes of organisms could structure newly assembled communities

22   that lack the influences of biogeographic filters, which affect the likelihood that a

23   species colonizes the site, and ecological sorting, in which some species exclude

24   others through interspecific interactions, such as competition (Chave, 2009) and

25   predation (Holt, 2009).  Here, we investigate orderliness in the interaction web of a

1     new, artificially assembled community of plants and their pollinators and seek its

2     causes in individual-based mechanisms.

4     Interactions between guilds of mutualists can be construed as 'two-node' networks

5     and depicted as bipartite graphs (Jordano, 1987). The networks have two kinds of

6     node with links that represent interactions only between nodes of different types. For

7     plant-pollinator communities, the two node types relate to the pollinators and the

8     plants. Plant-pollinator networks from a wide range of natural communities exhibit a

9     variety of recurrent statistical properties (Vásquez *et al.*, 2009), one of which is

10     'nestedness' (Bascompte *et al.*, 2003; Patterson and Atmar, 1986). A mutualist

11     network exhibits nestedness when nodes vary in the number of partners they have

12     and the partners of any one node are a perfect subset of the partners of other nodes

13     of the same type. Nestedness implies that nodes in a particular layer with few

14     connections, or specialists, are connected to nodes in the other layer that are

15     themselves connected to highly connected generalists in the original layer. In effect,

16     this means that direct interactions between specialists are absent or rare and species

17     of the same guild interact indirectly through generalists. Strong nestedness therefore

18     indicates a high degree of organization in the community. As noted above, it is

19     natural to enquire about the causes of this striking structural motif and two ecological

20     mechanisms have been implicated as causes of nestedness in plant-pollinator

21     networks; 'interaction neutrality' (Krishna *et al.*, 2008), which occurs when the

22     probability of interaction between two species is the product of their relative

23     abundances; and 'size matching' (Stang *et al.*, 2006; Stang *et al.*, 2007), which

24     occurs when the proboscis length of a nectivorous pollinator species must exceed a

25     threshold to exploit the nectar of a particular plant species – the threshold being

1  determined by the depth of the nectary in that species' flowers.  Here, we investigate

2  nestedness in a community where these two mechanisms are particularly weak.

4  The observations that we analyze were made on a collection of milkvetches

5  (*Astragalus* spp.) and sainfoins (*Onobrychis* spp.) in a botanical garden in the

6  northwest of the United States of America (Clement *et al.*, 2006).  We term these

7  plants and their associated bee fauna the 'Pullman community' and we refer to the

8  associated network as the 'Pullman network'.  Morphologically, the flowers of the

9  Pullman community are exceptionally homogeneous compared to those in a natural

10  community, because the species are distributed only between closely related clades

11  in the legume family (Wojciechowski *et al.*, 2004).  Collectively, flowers vary among

12  these species in size, colour and arrangement on inflorescences, but their basic

13  geometry is remarkably constant (Herrera, 2005), because their flowers are simply

14  morphological variations on the papilionaceous (pea-flower) groundplan - a showy

15  banner petal, two keel petals that surround the joined stamens and pistil, and two

16  wing petals (Arroyo, 1981).  Even though the *Astragalus* clade contains ten times as

17  many species as *Onobrychis*' clade (it is among the most species-rich genera of the

18  angiosperms, with approximately 2500 species), there is no distinctive morphological

19  innovation, floral or vegetative, to explain this (Sanderson and Wojciechowski, 1996).

20  Consequently, size matching originating in variation in floral morphology is unlikely to

21  be a strong cause of network nestedness.  Interaction neutrality is also likely to be

22  weak, because the relative abundance of plant species was equal by virtue of the

23  constant size of the nursery beds.  Why, then, should we expect the Pullman network

24  to be nested?

1     We propose a mechanism arising from the evolution of a pollen diet among bees.

2     The typical pollinator fauna of milkvetches and sainfoins in North America comprised

3     two kinds of social bees - the introduced honeybee (*Apis mellifera* L.) and native

4     bumblebees (*Bombus* spp.) - and various native solitary bees, such as digger bees

5     (*Anthophora* spp.), mason bees (*Osmia* spp.) and sweat bees (*Halictidae* spp.).

6     These bees land on the keel petals and thereby trip the flower's mechanism to reach

7     the nectar and pollen (Green and Bohart, 1975; Karron, 1987). The bee species in

8     the Pullman community have various levels of dietary specialization. Bees are

9     termed 'polylectic' if they collect pollen from a wide variety of flowers and 'oligolectic'

10     if the range of their pollen diet is narrow (Cane and Sipes, 2006). These dietary

11     variations can cause network nestedness as follows. Oligolecty is the likely ancestral

12     state in bees (Larkin *et al.*, 2008; Sedivy *et al.*, 2008) and polylecty is a derived trait

13     that emerges when a lineage develops the capacity to overcome certain 'constraints',

14     which are postulated to arise in pollen digestion and neurological limits on the

15     recognition and handling of flowers (Sedivy *et al.*, 2008). If the evolution of polylecty

16     merely extends the ancestral oligolectic diet, because lineages retain the ability to

17     use pollen from an ancestral host-plant (Larkin *et al.*, 2008), the ancestral oligolectic

18     diets will be a subset of the derived polylectic diets and nestedness will result. As

19     observed (Jordano *et al.*, 2006), nestedness will be widespread among plant-

20     pollinator communities, because oligolectic and polylectic species coexist in virtually

21     all bee faunas so far investigated (Sedivy *et al.*, 2008). If the plant species in the

22     Pullman community vary in pollen quality for different bees, and bees can

23     discriminate among their flowers based on this, we should expect nestedness in the

24     Pullman network. If we are correct in proposing that the overall similarity among

25     flowers in the Pullman community means that mechanical constraints (the feasibility

1   of visiting a flower) and neurological constraints (the feasibility of recognizing and

2   handling a flower) do not limit the dietary range of bees, an evaluation of nestedness

3   among pollinators is a test of the capacity for the biochemical attributes of pollen to

4   structure a plant-pollinator community.

6   Specialized pollination in plants is typically considered as being achieved through

7   morphological adaptations of the flower, such as long spurs that act as deep wells to

8   conceal nectar from all except the longest-tongued visitors, and these adaptations

9   function as 'filters' to exclude particular floral visitors (Johnson and Steiner, 2000).  If

10  these filters are phylogenetically derived traits, then the 'pollinator diets' of

11  specialized plants will be subsets of an ancestral generalized diet, which could cause

12  nestedness.  However, nestedness may arise despite morphological homogeneity

13  among flowers, if there are derived biochemical traits that also act as filters, such as

14  floral scent (Dotterl and Vereecken, 2010) and nectar allelochemicals (Shuttleworth

15  and Johnson, 2009).   Potentially, the biochemical composition of pollen could be a

16  filter and its evolution could contribute to nestedness in the Pullman network.  If we

17  are correct in proposing that morphological filters against bees are largely absent

18  among the flowers of the Pullman network, an evaluation of nestedness among

19  plants is a test of the capacity for biochemical attributes of pollen to structure a plant-

20  pollinator community.

22  The mechanism for generating nestedness that we propose has a phylogenetic

23  basis.  If it operates on the Pullman network, we expect to detect strong phylogenetic

24  influence in other ways.  We therefore evaluate the Pullman network for phylogenetic

25  signal, which is the tendency of evolutionarily-related species to resemble each other

1  more than species drawn at random from across the evolutionary tree (Blomberg and

2  Garland, 2002).  In mutualist networks, the presence of phylogenetic signal means

3  that related species will tend to interact with the same subset of partner mutualists.

4  Phylogenetic signal is a recurring pattern in natural plant-pollinator networks

5  (Rezende *et al.*, 2007) and we therefore test for it in our artificial community.  We

6  also test for a further phylogenetic influence on network structure, 'phylogenetic

7  reciprocy', which we define as the tendency of the partners of evolutionarily related

8  species to be, themselves, related.

11                          MATERIALS AND METHODS

13  *Study system*

14  The plant-pollinator community that we analysed was described by Clement *et al*.

15  (2006).  It was situated in two adjoining open field nurseries of the USDA's Western

16  Regional Plant Introduction Station in Pullman, Washington, USA (46°40' 13'' N 117°

17  45' 8'' W) among steppe vegetation comprising a mosaic of grasses, balsam root

18  (*Balsamorhiza sagitta* Pursh.; Asteraceae), lupines (*Lupinus* spp.; Fabaceae) and

19  various shrubs, such hawthorne (*Crataegus* spp.; Rosaceae) and rose (*Rosa* spp.;

20  Rosaceae).   Each nursery contained a collection of species from a single genus: the

21  *Astragalus* nursery comprised 68 species and the *Onobrychis* nursery comprised 35

22  species.  Individual plants were spaced 0.4 m apart in rows spaced 1.5 m apart.

23  Individuals from each plant species occupied adjacent positions.  Bees in the

24  nurseries were censused on 18 occasions during 1989-93, which occurred between

25  May and June of each year.  Selected plant species comprised between three and 18

1   flowering individuals and each was sampled for 15 minutes each day between each

2   between 11:00 and 15:00 by using insect nets to capture bees for subsequent

3   identification.  Bees were collected both on flowers and while patrolling above and

4   around flowers.

6   To test the adequacy of the procedure for collecting bees, we examined whether

7   there was an asymptote in the species accumulation curve, i.e. the relationship

8   between the number of bee species collected and the number of plant species

9   sampled.   We fitted a three-parameter asymptotic exponential curve to this

10  relationship (i.e. number of bee species collected = $a - be^{-c \times plants}$, where $a$ and b are

11  fitted constants and *plants* is the number of plant species from which observations of

12  bees are included).  To evaluate the sampling variability in the asymptotic values of

13  the number of bee species collected, we fitted this curve for each of 20 random

14  permutations of the order in which plant species were observed.

16  *Network nestedness*

17  We constructed a bipartite network for the Pullman community by linking plant

18  species to their floral visitors and we assumed that patrolling behaviour indicated of

19  floral utilization, so that all captured insects were taken to be flower visitors.  To

20  quantify the Pullman network's level of nestedness, various statistical indices can be

21  used (Ulrich *et al.*, 2009), but most are so-called 'distance metrics' that evaluate the

22  nestedness of the entire network.  Instead, we used an index that can quantify

23  separately the nestedness of each guild of mutualists (Burgos *et al.*, 2009).  The

24  index is based on the Attack Tolerance Curve (ATC; Memmott *et al.*, 2004), which

25  indicates the response of a network to the elimination of a given fraction of species of

1     one guild, after which some species of the other guild may be left without connection

2     and hence become extinct.  Thus, the ATC relates the fraction of surviving species of

3     one guild to the fraction of eliminated (attacked) species of the other guild.  The ATC

4     of a network depends on its level of nestedness and the attack strategy deployed

5     against it.  If the attack strategy eliminates species in one guild by beginning with the

6     most highly connected and proceeding rankwise to the least connected, then

7     perfectly nested networks have a distinctive, convex-shaped ATC signature.  If,

8     instead, elimination proceeds from the least connected, the ATC signature is

9     concave.  The quantitative difference between these two signatures can be used as

10     an index of nestedness (Fig. 1; Burgos et al. 2009), as follows.

12     Let the robustness coefficient, $R_A$, be defined as the area under the ATC, which

13     depends on the particular attack strategy, denoted by the subscript $A$. There are two

14     alternative, extreme attack strategies: eliminate nodes of a given type rankwise

15     starting from the highest degree, denoted $A = (+ \rightarrow -)$; and the opposite scheme,

16     starting from the lowest degree, denoted $A = (- \rightarrow +)$, where a node's degree is the

17     number of connections to it.  In the case of perfect nestedness, the attack strategy $(-$

18     $\rightarrow +)$ yields $R_{(- \rightarrow +)} = 1$ and $(+ \rightarrow -)$ yields $R_{(+ \rightarrow -)} = \phi$.  We define the nestedness

19     coefficient, $H$, as the normalized difference between these robustness coefficients,

20     i.e.

22     $H = (R_{(- \rightarrow +)} - R_{(+ \rightarrow -)}) / (1 - \phi)$.                          (Eq. 1)

24     This coefficient is correctly bounded, because $H = 1$ for perfect nestedness and $H$

25     decreases for increasingly random networks, with $H = 0$ in the perfectly anti-nested

1   case.  We determined the sampling distribution of $H$ under the null hypothesis that

2   species interactions were allocated among nodes randomly and independently, but

3   with the proviso that the degree distributions are fixed in the attacked guild.

4   Biologically, we compare the observed nestedness of a guild to a randomized version

5   of itself, which is one that is connected at random to the available mutualist partners

6   who themselves have pre-defined diet breadths.  Statistically, this makes our

7   significance tests more conservative than in a comparison between the observed

8   network structure and the collection of networks with completely randomized

9   connections (Ulrich et al.).  These Monte Carlo randomizations were implemented in

10  Fortran95 (Koelbel *et al.*, 1994).  For comparison with previous analyses (Jordano et

11  al. 2006), we also calculated a standard index of 'nestedness temperature', denoted

12  $T$ (Atmar and Patterson, 1995) and report the nestedness coefficient, $N$, which is

13  calculated as $N = (100 - T)/100$.

16  *Phylogenetic signal*

17  Initially, we tested whether the clades of each guild were ecologically differentiated

18  by Analysis of Molecular Variance (AMOVA) as follows.  We summarized the

19  observed pairwise direct interactions in the community as an adjacency matrix, **K**,

20  where the list of interactions for each species of one guild occupies a row and each

21  of the columns corresponds to a species of the other guild.  Each element of **K** is

22  identified by its location in the $i^{th}$ row and $j^{th}$ column and it takes the value of unity if

23  the $i^{th}$ species of one guild interacts with the  $j^{th}$ species of the other guild, and zero

24  otherwise. For each species, the details of its interactions are a list of zeros and

25  ones, which is formally homologous to the binary haplotype data that is analysed by

1     population geneticists using AMOVA. Rather than seeking differentiation in

2     haplotypes among populations, we are seeking differentiation in interaction profiles

3     among clades, which were either genera (plants) or families (bees). We

4     implemented AMOVA and the PCA ordination in GenAlEx v. 6.41(Peakall and

5     Smouse 2006).

7     If significant phylogenetic differentiation was detected, we then compared clades

8     within each guild pairwise as follows. First, we described the 'ecological similarity'

9     between each pair of species, $a$ and $b$, by quantifying the level of similarity in their

10     array of partners using the simple matching coefficient, $S_{a,b} = M/C$, where $M$ is the

11     number of matching partners, i.e. partners that visit both $a$ and $b$, and $C$ is the

12     number of partners compared, i.e. the set of partners of $a$ and $b$. Consider a pair of

13     clades, denoted A and B, that comprise sets of species $\{a_1, a_2, \ldots a_m\}$ and $\{b_1, b_2, \ldots$

14     $b_n,\}$ respectively. Within each clade, we calculated the mean ecological similarity of

15     all possible pairs of species, denoted $<S_{A,A}>$ and $<S_{B,B}>$, and calculated their grand

16     mean $<S^*_{within}>$. We also calculated the ecological similarity between all possible

17     pairs of species drawn one per clade and calculated their grand mean $<S^*_{between}>$.

18     By analogy to the $F$ statistic in Analysis of Variance, we calculated the ratio $F^* =$

19     $<S^*_{within}>/<S^*_{between}>$, which takes a value greater than unity if the two clades are

20     differentiated in their mutualist partners. We tested the statistical significance of

21     deviations of $F^*$ from unity by determining the critical values of the sampling

22     distribution of $F^*$ under the null hypothesis that the composition of a species

23     interactions was independent of phylogeny. We pooled the $(m + n)$ species of clades

24     $A$ and $B$ and used a Monte Carlo randomization to assign them to two groups of size

25     $m$ and $n$, respectively, and calculated $F^*$ as described above. After at least 1000

1    iterations of this randomization, the 95[th] percentile of the resulting sampling

2    distribution of $F*$ served as the critical value for rejecting the null hypothesis with a

3    confidence level of P<0.05, etc.  Rather than applying this procedure to all possible

4    pairs of clades within a guild, which would inflate the rate of Type 1 error, we

5    proceeded in the style of an orthogonal contrast and compared each clade only with

6    its sister.  These Monte Carlo randomizations were implemented in **R** (Ihaka and

7    Gentleman, 1996).

10   *Phylogenetic reciprocy*

12   We tested whether the Pullman network contained significant phylogenetic reciprocy

13   as follows.  We defined the phylogenetic distance between two species in the same

14   mutualist guild as the number of cladistic bifurcations that are passed in travelling the

15   shortest route along the phylogenetic tree from one species to the other, with the

16   proviso that the total distance between any species and the root of the tree is equal,

17   i.e. that the distances are ultrametric.  Thus, even if a species has no sister taxa

18   between itself and the root of the phylogeny, the phylogenetic distance between that

19   species and the root of the tree is taken to be equal to the greatest number of

20   bifurcations between any species and the root of the tree, etc., which in effect credits

21   both phyletic and cladistic evolutionary change in our quantification of phylogenetic

22   distance.  For a given clade of one mutualist guild (genus for plants, family for bees),

23   we calculated the mean phylogenetic distance among its mutualist partners given the

24   observed topology of the network and of the phylogenetic tree, denoted $<d_{obs}>$, as

25   follows.

2 Let $\mathbf{K'}$ denote an adjacency matrix whose rows are a subset of those of the

3 community's adjacency matrix, $\mathbf{K}$, such that it describes the mutualist network only

4 for those species that belong to a designated clade. Let $\mathbf{D}$ denote a lower

5 triangular matrix, whose non-zero entries are each a phylogenetic distance among

6 pairs of species of the other mutualist guild. The phylogenetic distances among the

7 pairs of members of this guild linked through members of the other guild are the

8 elements of $\mathbf{W}$, a matrix that is given by the product $\mathbf{K_T}\mathbf{K}\mathbf{D}$, where $\mathbf{K_T}$ denotes the

9 transpose of $\mathbf{K}$. The mean phylogenetic distance among the mutualist partners

10 given the observed topology of the network and of the phylogenetic tree, $<d_{obs}>$, is

11 given by:

13 $$<d_{obs}> = \sum_{i,j} \mathbf{W} / k_{total} \qquad\qquad (\text{Eq 2})$$

14 where $k_{total}$ is the number of mutualist pairs that are connected through the network

15 specified by $\mathbf{K'}$. To determine the sampling distribution of $<d>$ under the null

16 hypothesis that a clade interacts with its mutualists independently of their

17 phylogenetic relatedness, we randomized the mutualist's phylogeny in relation to the

18 network by permuting the labels of the rows and columns of $\mathbf{D}$, which maintains the

19 topology of the phylogeny, but randomly connects it to the network . For each

20 randomization, we used Eq 2 to calculate $<d_{rand}>$, the mean phylogenetic distance

21 among the connected mutualists given the observed topology of the network and the

22 randomized phylogenetic tree. The Monte Carlo sampling distribution of $<d_{rand}>$ was

23 then obtained from at least 1000 permutations of the phylogenetic tree and its critical

24 percentiles compared with $<d_{obs}>$.

1   For bees, we used a published phylogeny (Danforth *et al.*, 2006) with the following

2   taxa including the species at the branch tips: Andrenidae, non-social Apidae

3   (Anthophora, Eucera), Social Apidae (*A. melifera* and *Bombus* spp.), Colletidae,

4   Halictidae, and Megachilidae, which yielded a maximum ultrametric pairwise distance

5   between species of $d = 10$. For plants, we used a simple, unresolved tree with two

6   taxa, *Astragalus* spp. and *Onobrychis* spp., that included the species at the branch

7   tips, which yielded a maximum ultrametric distance of $d = 4$. These Monte Carlo

8   randomizations were implemented in **R** (Ihaka and Gentleman, 1996).

11                              RESULTS

13   The field investigators collected 52 bee species that were associated with 37 plant

14   species. For bees, the fitted species accumulation curves were well described by

15   three-parameter asymptotic exponential relationships (mean r-squared = 0.97, SD =

16   0.016, n = 20) and these curves had an average asymptotic value of approximately

17   56.5 (SE = 1.94, n = 19). The observed species richness was significantly lower than

18   this (*t*-test, $t = 2.35$, df = 19, P < 0.05). These results indicate that the collections

19   from the nurseries had sampled almost all of the available bee species in the locality,

20   but that we estimate that about four bee species were missing from the network.

22   The community yielded 258 pairwise interactions (Fig. 2) from a potential of ($52 \times 37$)

23   = 1924, which is a network density of 13.4%. Among bees, the mean number of

24   plant partners was 5.0 (SE = 0.70, n = 52; Fig.1) and the mean pairwise ecological

25   similarity was <S> = 0.12 (SE = 0.004, n = 2704). Among plants, the mean number of

1  insect partners was 7.0 (SE = 0.90, n = 37; Fig. 1) and the mean pairwise ecological

2  similarity was $<S>$ = 0.14 (SE = 0.004, n = 1332).

4  *Nestedness*

5  According to the conventional index, the network is significantly nested (Fig. 3; $N$ =

6  0.88, P < 0.05).  For both bees and plants, the Pullman network is significantly more

7  nested than randomized networks ($H \approx 0.55$, P < 0.001; Fig. 4), but the ATCs are

8  straight for both animals and plants, which implies that the system is at the order-

9  disorder transition.

11  *Phylogenetic signal*

12  Overall, the clades of bees were significantly differentiated in floral associations

13  (AMOVA, PhiPT=0.08, P = 0.013) and there was significant pairwise differentiation

14  among some sister clades (Fig. 5).  The two genera of plants were significantly

15  differentiated in their mutualistic partners; the mean pairwise ecological similarity

16  within genera was $<S_{within}>$ = 0.17 and the ratio of similarity between vs. within the

17  genera was $F^*$ = 1.68 (P<0.001).

19  *Phylogenetic reciprocy*

20  Phylogenetic reciprocy was evident among clades of both plant and animal

21  mutualists.  The partners of various clades of bees were themselves significantly

22  closely related (Fig. 6a).  This occurred when clades of bees had dietary preferences

23  for species of *Astragalus* (e.g. Megachilidae, Apidae).  Similarly, the partners of the

24  plant genera were significantly closely related (Fig. 6b).

3                                    DISCUSSION

5    Despite its artificial origin, the Pullman network has some attributes in the normal

6    range for networks describing natural plant-pollinator communities, such its density

7    and degree distributions, but with respect to nestedness, it is special.  The Pullman

8    network shows significant orderliness in this aspect, although networks from natural

9    systems are more strongly nested, particularly for the animal guild where $0.6 \leq H \leq$

10   $0.75$ (Burgos $et\ al.$, 2009).  If we were to use a self-ordering algorithm to progress

11   step-wise towards a nested network starting from a random one, the network would

12   pass through an order-disorder transition, which is situated at the point where the

13   ATC curves are flat and $H = 0.5$ (Burgos $et\ al.$, 2009).  The Pullman network is

14   almost precisely at this transition.  Whatever factors are shaping the Pullman

15   network, we may say that they have brought it to a non-random state of incipient

16   nestedness.  What are the factors responsible?

18   We have argued that neither interaction neutrality nor size matching are likely to be

19   strong causes of orderliness.  The evenness in the abundances of plants in the

20   nursery beds that was intrinsic to the study's design undermines the influence of

21   interaction neutrality and, furthermore, certain results yield tangible evidence to

22   support our claim that size matching is also weak.  Specifically, all but two of the

23   plant species were visited by at least one species of solitary bee.  Solitary bees are

24   generally smaller than social bees, but it was apparently feasible for a small bee to

25   enter and exploit virtually any of the flowers in the Pullman community.

1     Consequently, we argue that there was no minimum size threshold that restricted

2     certain plants to only the larger insects.   We have instead proposed that the

3     evolution of pollen diets in bee lineages (Sedivy *et al.*, 2008) could be a source of

4     nestedness, but what support for this 'dietary constraint hypothesis' is provided by

5     our analysis of the Pullman network?

7     When considered at the level of taxonomic family, many of the bee lineages

8     interacted with statistically distinct sets of plant partners, which is consistent with the

9     proposition that lineages experience dietary constraints.  These sets were not sharply

10     defined, however, because the levels of pairwise dietary similarity among bee

11     species within lineages was relatively low, being only about 1.4 times greater than

12     the background pairwise similarity of c. 10% plant species in common.  Furthermore,

13     we also found that the mutualist partners of bee lineages were themselves related,

14     which we termed phylogenetic reciprocy.  When applied to families of bees with the

15     more narrow diets, this finding is a quantitative demonstration of the phenomenon of

16     oligolecty in the strictest sense, namely that the plants that comprise an oligolectic

17     diet shall be themselves taxonomically restricted (Cane and Sipes, 2006).  When this

18     phylogenetic relatedness among plants is reflected in the biochemical similarity of

19     pollen (e.g. Weiner *et al.*, 2010), this too is consistent with the proposition that dietary

20     constraints limit the taxonomic scope of bee lineages for establishing partners.  We

21     therefore postulate that biochemical constraints on bee diets make the network

22     orderly, but that they are only sufficiently powerful to bring the Pullman network to

23     incipient nestedness.  Presumably, natural communities are more strongly nested

24     because of the additional contributions of size matching and interactive neutrality .

1    *Caveats and prospects*

3    Our description of the Pullman network is probably be missing a few bee species and

4    contains some inappropriate links, because patrolling bees were not distinguished

5    from flower visitors.  However, we do not think that these inaccuracies bias our

6    analysis of nestedness (Araujo *et al.*, 2010; Vazquez *et al.*, 2009) and the pervasive

7    phylogenetic orderliness that we found in the network gives us confidence that it has

8    abstracted important aspects of biological reality.

10   We recognize that the structure of the Pullman network provides only weak support

11   for the hypothesis that its incipient nestedness originated in phylogenetically

12   arranged constraints on the pollen diets of insects, not least because it is possible to

13   invoke an alternative that fits the observed pattern equally well, namely that subtle

14   morphologically-based size matching had created the pattern.  Formally, there are

15   nine criteria for scientifically implicating a factor as the cause of a biological effect

16   (Hill, 1965) and our hypothesis that dietary constraint causes nestedness meets only

17   three.  Specifically, we have proposed a plausible mechanism, it is analogous to

18   other known mechanisms (e.g. the power of secondary chemicals to organize plant-

19   herbivore communities; Becerra, 2007), and it is coherent with known facts about

20   dietary selectivity in bees and the differentials among plant species in the nutritive

21   value of their pollens to bees (Roulston and Cane, 2002; Tasei and Aupinel, 2008).

22   However, our explanation fails to meet other key criteria, such as the need to show

23   that similar effects on nestedness occur consistently in a range of studies and that

24   the effect is demonstrated experimentally.  However, these criteria help to identify

25   some requirements for further work, as follows.  First, it will be important to

1  investigate nestedness in networks from additional communities where the

2  contribution of other mechanisms (e.g. size matching) is controlled, which is probably

3  most feasible in artificial plant communities.  Second, besides the basic phenomenon

4  of oligolecty, some further particularities of the 'dietary constraint' hypothesis should

5  be investigated mechanistically.  Specifically, its validity depends on the existence of

6  nestedness in the biochemical profiles of plant pollens and, furthermore, on the ability

7  of pollinators to make flower choices based on recognising this level of variation,

8  perhaps through associated floral volatiles (Raguso, 2008).  In essence, we propose

9  that valuable progress in understanding plant-pollinator systems may be made

10  through applying the techniques of chemical ecology at the community scale, as has

11  been done elsewhere (Becerra, 2007).

14                    ACKNOWLEDGEMENTS

16  L.H. is grateful for enlightening discussions with H.Ceva, E.Burgos and R.P.Perazzo

17  concerning nestedness and its measure.  J.E.C. respectfully dedicates this work to

18  Beverly Rathcke, his PhD supervisor.

LITERATURE CITED

**Araujo AIL, de Almeida AM, Cardoso MZ, Corso G. 2010.** Abundance and nestedness in interaction networks. *Ecological Complexity* **7**:494-499.

**Arroyo M. 1981**. Breeding systems and pollination biology in Leguminosae. In: Polhill R, Raven P (eds) *Advances in Legume systematics, Part 2*. London: Royal Botanic Gardens, Kew.

**Atmar W, Patterson BD. 1995.** *The nestedness temperature calculator: a visual basic program, including 294 presence-absence matrices*. New Mexico: AICS Research, Inc.

**Bascompte J, Jordano P, Melian C, Olesen J. 2003.** The nested assembly of plant-animal mutualistic networks. *Proceedings of the National Academy of Sciences of the United States of America* **100**:9383-9387.

**Becerra JX. 2007.** The impact of herbivore-plant coevolution on plant community structure. *Proceedings of the National Academy of Sciences of the United States of America* **104**:7483-7488.

**Beckerman A, Petchey O, Warren P. 2006**. Foraging biology predicts food web complexity. *Proceedings of the National Academy of Sciences of the United States of America* **103**:13745-13749.

1  **Blomberg SP, Garland T. 2002.** Tempo and mode in evolution: phylogenetic inertia,

2      adaptation and comparative methods. *Journal of Evolutionary Biology* **15**:899-

3      910.

5  **Briand F, Cohen J. 1984.** Community food webs have scale-invariant structure.

6      *Nature* **307**:264-267.

8  **Burgos E, Ceva H, Hernandez L, Perazzo RPJ. 2009.** Understanding and

9      characterizing nestedness in mutualistic bipartite networks. *Computer Physics*

10      *Communications* **180**:532-535.

12  **Cane J, Sipes S. 2006.** Characterizing floral specialization in bees: analytical

13      methods and a revised lexicon for oligolecty. In: Waser N, Ollerton J (eds)

14      *Plant-pollinator interacti*ons. Chicago: University of Chicago Press, 99-122.

16  **Chave J. 2009.** Competition, neutrality, and community organization. In: Levin S (ed)

17      *The Princeton guide to ecology*. Princeton: Princeton University Press.

19  **Clement S, Griswold T, Rust R, Hellier B, Stout D. 2006.** Bee associates of

20      flowering *Astragalus* and *Onobrychis* genebank accessions at a Snake River

21      site in Eastern Washington. *Journal of the Kansas Entomological Society*

22      **79**:254-260.

1   **Cohen JE. 1989.** Food webs and community structure. In: Roughgarden J, May RM,
2       Levin SA (eds) *Perspectives in ecological theory*. Princeton, NJ: Princeton
3       University Press, 181-202.
4
5   **Connor E, Simberloff D. 1979.** The assembly of species communities: chance or
6       competition? *American Naturalist* **113**:791-833.
7
8   **Danforth BN, Sipes SD, Fang J, Brady SG. 2006.** The history of early bee
9       diversification based on five genes plus morphology. *Proceedings of the*
10      *National Academy of Sciences of the United States of America* **103**: 15118-15123
11
12  **Dotterl S, Vereecken NJ. 2010.** The chemical ecology and evolution of bee-flower
13      interactions: a review and perspectives. *Canadian Journal of Zoology* **88**:668-
14      697.
15
16  **Dunne J. 2005.** The network structure of food webs. In: Pascual M, Dunne J (eds)
17      *Ecological networks*. Oxford: Oxford University Press.
18
19  **Elton C. 1927.** *Animal ecology*. London: Sidgwick and Jackson, Ltd.
20
21  **Green T, Bohart G. 1975.** The pollination ecology of *Astragalus cibarius* and
22      *Astragalus utahensis* (Leguminosae). *American Journal of Botany* **62**:379-386.
23
24  **Herrera J. 2005.** Phenotypic correlations among plant parts in Iberian Papilionoideae
25      (Fabaceae). *Annals Of Botany* **95**:345-350.

**Hill A. 1965.** The environment and disease: association or causation? *Proceedings of the Royal Society of Medicine* **58**:295-300.

**Holt R. 2009**. Predation and community organization. In: Levin S (ed) *The Princeton guide to ecology.* Princeton, NJ: Princeton University Press.

**Hutchinson G. 1953.** The concept of pattern in ecology. *Proceedings of the Academy of Natural Sciences of Philadelphia* **105**:1-12.

**Ihaka R, Gentleman R. 1996.** A language for data analysis and graphics. *Journal of Computational and Graphical Statistics* **5**:299-314.

**Johnson SD, Steiner KE. 2000.** Generalization versus specialization in plant pollination systems. *Trends in Ecology & Evolution* **15**:140-143.

**Jordano P. 1987.** Patterns of mutualistic interactions in pollination and seed dispersal: connectance, dependence asymmetries, and coevolution. *American Naturalist* **129**:657-677.

**Jordano P, Bascompte J, Olesen J. 2006.** The ecological consequences of complex topology and nested structure in pollination webs. In: Waser N, Ollerton J (eds) *Plant-pollinator interactions.* Chicago: University of Chicago Press, 173-199.

**Karron J. 1987.** The pollination ecology of co-occuring geographically restricted and widespread species of Astragalus (Fabaceae). *Biological Conservation* **39**:179-193.

**Koelbel C, Loveman D, Schreiber R, Steele G, Zosel M. 1994.** *The high performance Fortran handbook.* Cambridge, MA: MIT Press.

**Krishna A, Guimaraes P, Jordano P, Bascompte J. 2008.** A neutral-niche theory of nestedness in mutualistic networks. *Oikos* **117**:1609-1618.

**Larkin LL, Neff JL, Simpson BB. 2008.** The evolution of a pollen diet: Host choice and diet breadth of Andrena bees (Hymenoptera : Andrenidae). *Apidologie* **39**:133-145.

**Lindeman R. 1942.** The trophic-dynamic aspect of ecology. *Ecology* **23**:399-418.

**May R. 1972.** Will a large complex system be stable? *Nature* **238**:413-414.

**Memmott J, Waser NM, Price MV. 2004.** Tolerance of pollination networks to species extinctions. *Proceedings Of The Royal Society Of London Series B* **271**:2605-2611.

**Patterson B, Atmar W. 1986.** Nested subsets and the structure of insular mammalian faunas and archipelagos. *Biological Journal of the Linnean Society* **28**:65-82.

2  **Peakall R, Smouse P. 2006.** GENALEX 6: genetic analysis in Excel. *Molecular*

3      *Ecology Notes* **6**:288-295.

5  **Petchey O, Beckerman A, Riede J, Warren P. 2008.** Size, foraging, and food web

6      structure. *Proceedings of the National Academy of Sciences of the United*

7      *States of America* **105**:4191-4196.

9  **Raguso RA. 2008.** Wake Up and Smell the Roses: The ecology and evolution of

10     floral scent. *Annual Review of Ecology Evolution and Systematics* **39**:549-569.

12 **Rezende EL, Lavabre JE, Guimaraes PR, Jordano P, Bascompte J. 2007.** Non-

13     random coextinctions in phylogenetically structured mutualistic networks.

14     *Nature* **448**:925-926.

16 **Roulston TH, Cane JH. 2002.** The effect of pollen protein concentration on body

17     size in the sweat bee *Lasioglossum zephyrum* (Hymenoptera : Apiformes).

18     *Evolutionary Ecology* **16**:49-65.

20 **Sanderson M, Wojciechowski M. 1996.** Diversification rates in a temperate Legume

21     clade: are there "so many species" of *Astragalus* (Fabaceae)? *American*

22     *Journal of Botany* **83**:1488-1502.

1    **Sedivy C, Praz CJ, Muller A, Widmer A, Dorn S. 2008.** Patterns of host-plant

2         choice in bees of the genus *Chelostoma*: the constraint hypothesis of host-

3         range evolution in bees. *Evolution* **62**:2487-2507.

5    **Shuttleworth A, Johnson SD. 2009.** The importance of scent and nectar filters in a

6         specialized wasp-pollination system. *Functional Ecology* **23**:931-940.

8    **Stang M, Klinkhamer P, van der Meijden E. 2006.** Size constraints and flower

9         abundance determine the number of interactions in a plant–flower visitor web.

10         *Oikos* **112**:111–121.

12    **Stang M, Klinkhamer P, van der Meijden E. 2007.** Asymmetric specialization and

13         extinction risk in plant–flower visitor webs: a matter of morphology or

14         abundance? *Oecologia* **151**:442–453.

16    **Tasei JN, Aupinel P. 2008.** Nutritive value of 15 single pollens and pollen mixes

17         tested on larvae produced by bumblebee workers (*Bombus terrestris*,

18         Hymenoptera : Apidae). *Apidologie* **39**:397-409.

20    **Ulrich W, Almeida M, Gotelli NJ. 2009.** A consumer's guide to nestedness analysis.

21         *Oikos* **118**:3-17.

23    **Vásquez D, Blüthgen N, Cagnolo L, Chacoff N. 2009.** Uniting pattern and process

24         in plant–animal mutualistic networks: a review. *Annals Of Botany* **103**:1445-

25         1457.

**Vazquez DP, Chacoff NP, Cagnolo L. 2009.** Evaluating multiple determinants of the
structure of plant-animal mutualistic networks. *Ecology* **90**:2039-2046.

**Weiner CN, Hilpert A, Werner M, Linsenmair KE, Bluthgen N. 2010.** Pollen amino
acids and flower specialisation in solitary bees. *Apidologie* **41**:476-487.

**Wojciechowski M, Lavin M, Sanderson M. 2004.** A phylogeny of legumes
(Leguminosae) based on analysis of the plastid matK gene resolves many
well-supported subclades within the family. *American Journal of Botany*
**91**:1846-1862.

1    FIGURE LEGENDS

3    Fig. 1. Generic Attack Tolerance Curve (ATC) signatures for: (a) the strongly nested

4    network; (b) the randomly structured network; and (c) the network at an order-

5    disorder transition.  Open symbols indicate the ATC for the attack strategy that

6    proceeds stepwise from nodes of the lowest degree ($- \to +$) and closed symbols

7    indicate the ATC for ($+ \to -$).  Hatched areas indicate the magnitude of the numerator

8    of the coefficient of nestedness, $H$.  A thin line delineates an area under the ATC for

9    ($+ \to -$) in the perfectly nested network, i.e. $\phi$.

11    Fig 2.  The Pullman network, whose nodes are comprised of 52 species of bee and

12    37 species of *Astragalus* and *Onobrychis*.  Links indicate presumed plant-pollinator

13    interactions.  The length of the histogram bar above or below each node indicates the

14    number of links to that node, i.e. its degree, denoted by $k$.  The names of some taxa

15    are abbreviated as follows: Coll. = Colletidae; Andren. = Andrenidae; Antho. =

16    *Anthophora*; Am = *Apis mellifera*.  The species identity of each node is given in

17    Appendix 1.

19    Fig. 3. Sorted adjacency matrix of the Pullman community with each plant species

20    occupying a row, each bee species occupying a column, and plant-pollinator

21    interactions indicated by filled cells at appropriate row-column intersections.  The

22    curve approximates the isocline of maximal nestedness for a network with identical

23    degree distributions.

1    Fig 4.   The Attack Tolerance Curves (ATCs) for bee species, given the stepwise

2    elimination of plant species, and plant species, given the stepwise elimination of bee

3    species.   Open symbols indicate the ATC proceeding from nodes with the lowest

4    degree to the highest $(- \rightarrow +)$ and closed symbols indicate $(+ \rightarrow -)$.   The unadorned

5    curve indicates the average ATC of randomized networks, but where the attacked

6    guild retains the observed degree distribution.

8    Fig. 5.   The phylogenetic distribution of dietary differentiation between sister clades of

9    bees.   Asterisked figures at a cladistic bifurcation indicate significant differentiation

10    between the clades so marked, while a filled circle indicates a non-significant test

11    result.   Each figures at the junction of clades is the ratio of pairwise dietary similarity

12    between species within clades to the similarity between clades.   A single asterisk

13    indicates that the ratio is significantly greater than unity with $P < 0.05$, two asterisks

14    indicate $P < 0.01$.   The names of taxa are abbreviated as follows: Col = Colletidae;

15    Hal = Halictidae; And = Andrenidae; Meg = Megachilidae; Anth = solitary Apidae

16    (*Anthophora* and *Eucera*); and Soc = social Apidae (*A. mellifera* and *Bombus* spp.).

17    The mean pairwise dietary similarity between all pairs of bee species was *S* = 0.10.

19    Fig. 6. Phylogenetic reciprocy towards clades of bees (a) and plants (b).   An

20    asterisked figure superimposed on a lineage indicates that the mean phylogenetic

21    distance among pairs of the clade's mutualist partners is significantly lower than in

22    random networks, while a filled circle indicates a non-significant test result.   A single

23    asterisk indicates that the distance is significantly smaller with $P < 0.05$, two asterisks

24    indicate $P < 0.01$.   The names of taxa are abbreviated as follows: Col = Colletidae;

25    Hal = Halictidae; And = Andrenidae; Meg = Megachilidae; Anth = solitary Apidae

1    (*Anthophora* and *Eucera*); Soc = social Apidae (*A. mellifera* and *Bombus* spp.); Ast =

2    *Astragalus* spp.; and *Ono* = *Onobrychis* spp..  The mean phylogenetic distance

3    between pairs of species in random networks is 2.94 for partners of bees and 7.34

4    for partners of plants.

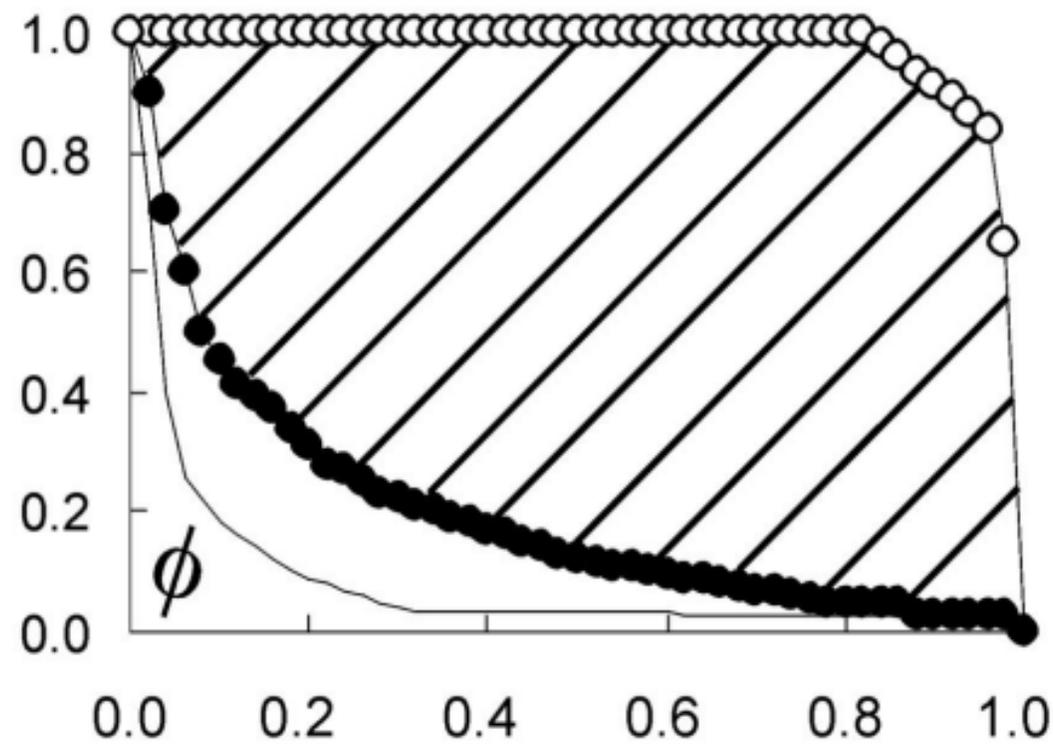
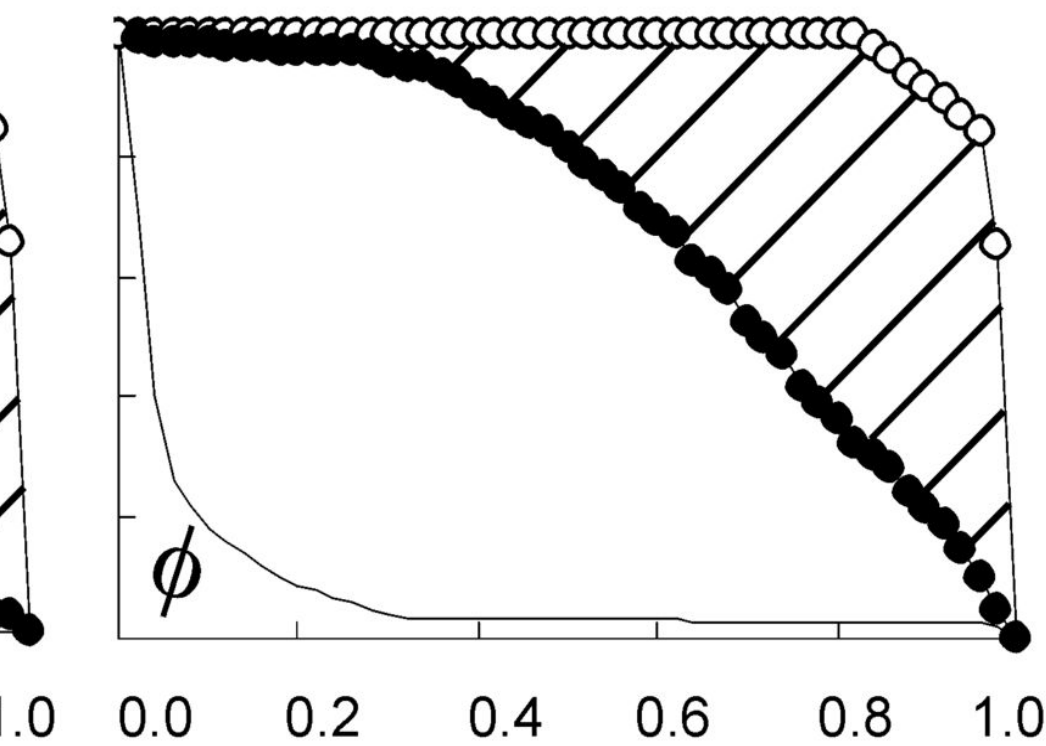
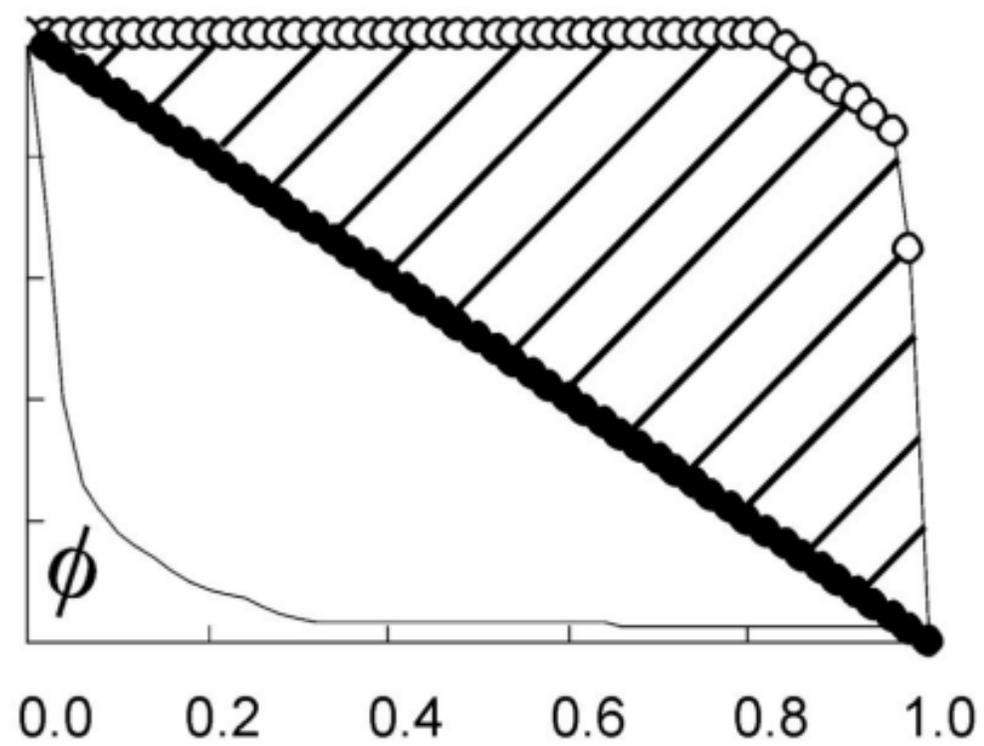

a.  b.  c.

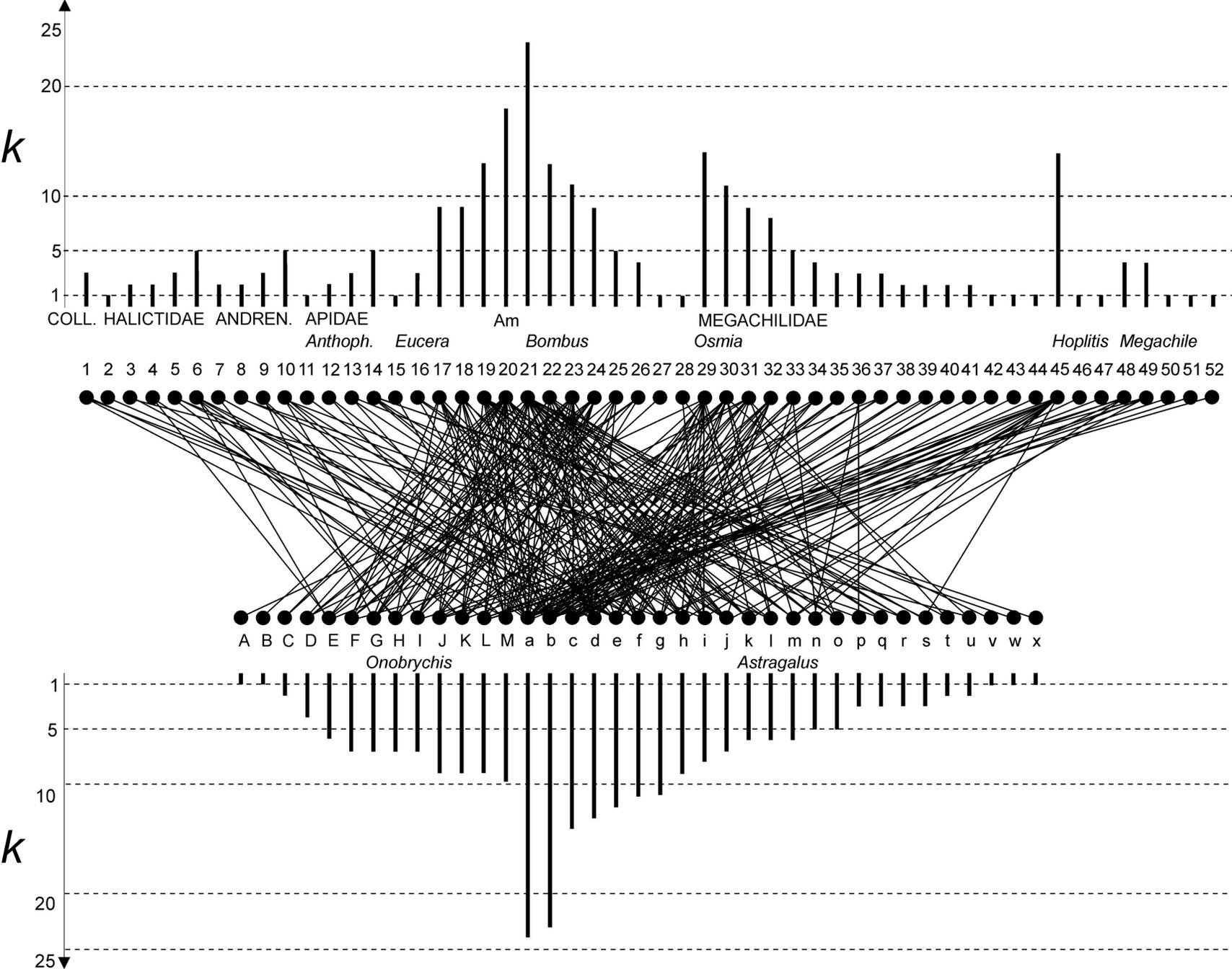

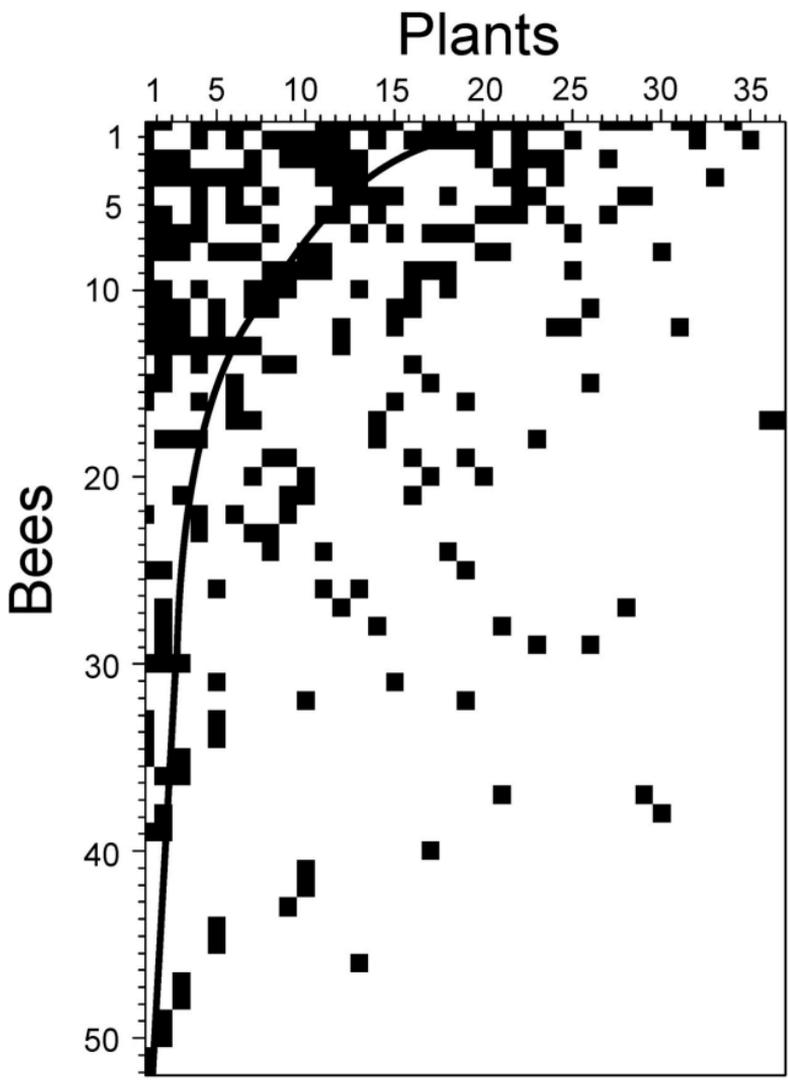

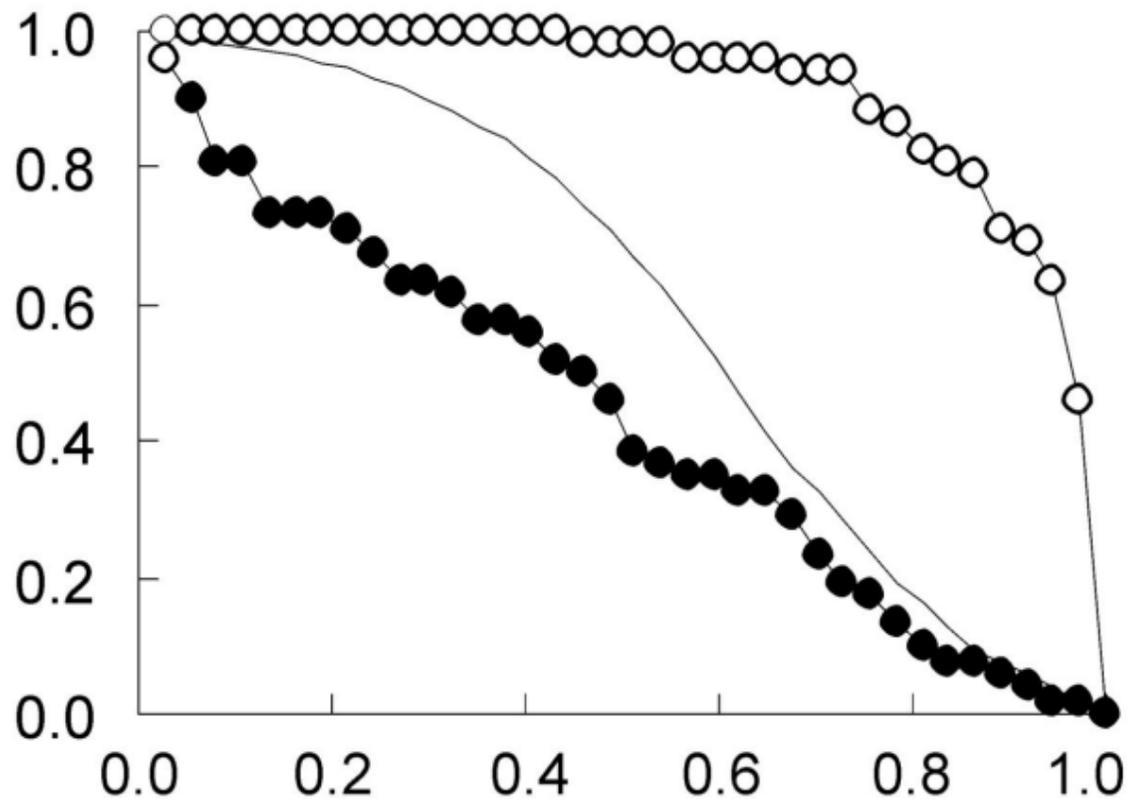
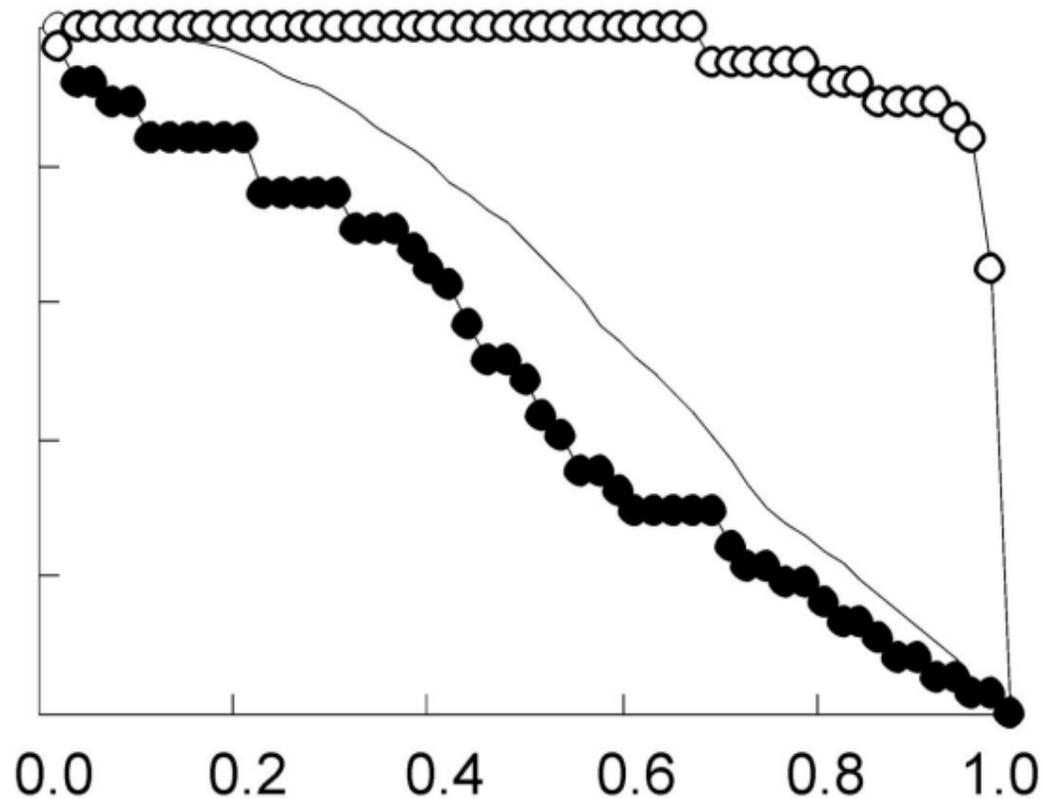

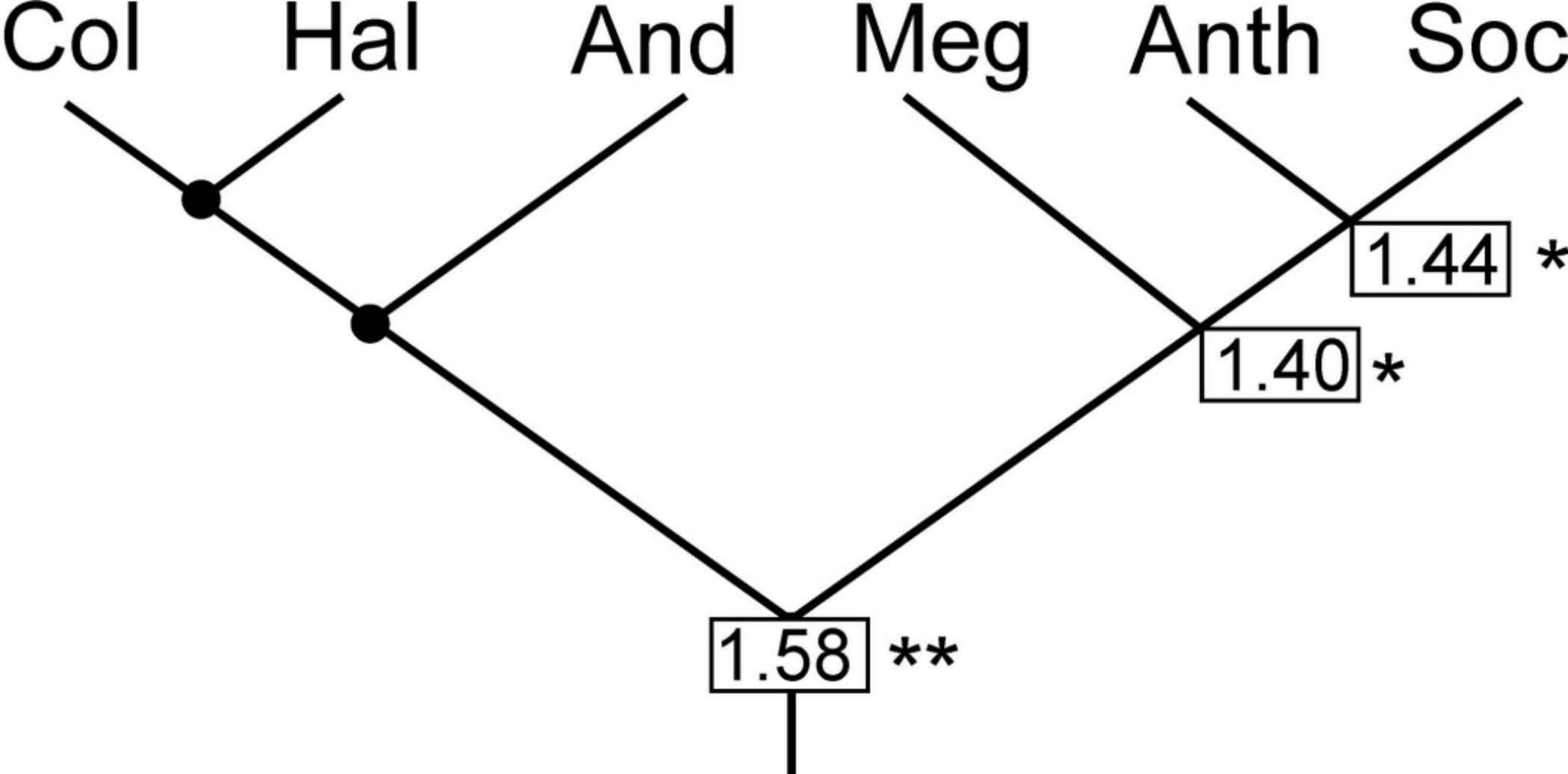

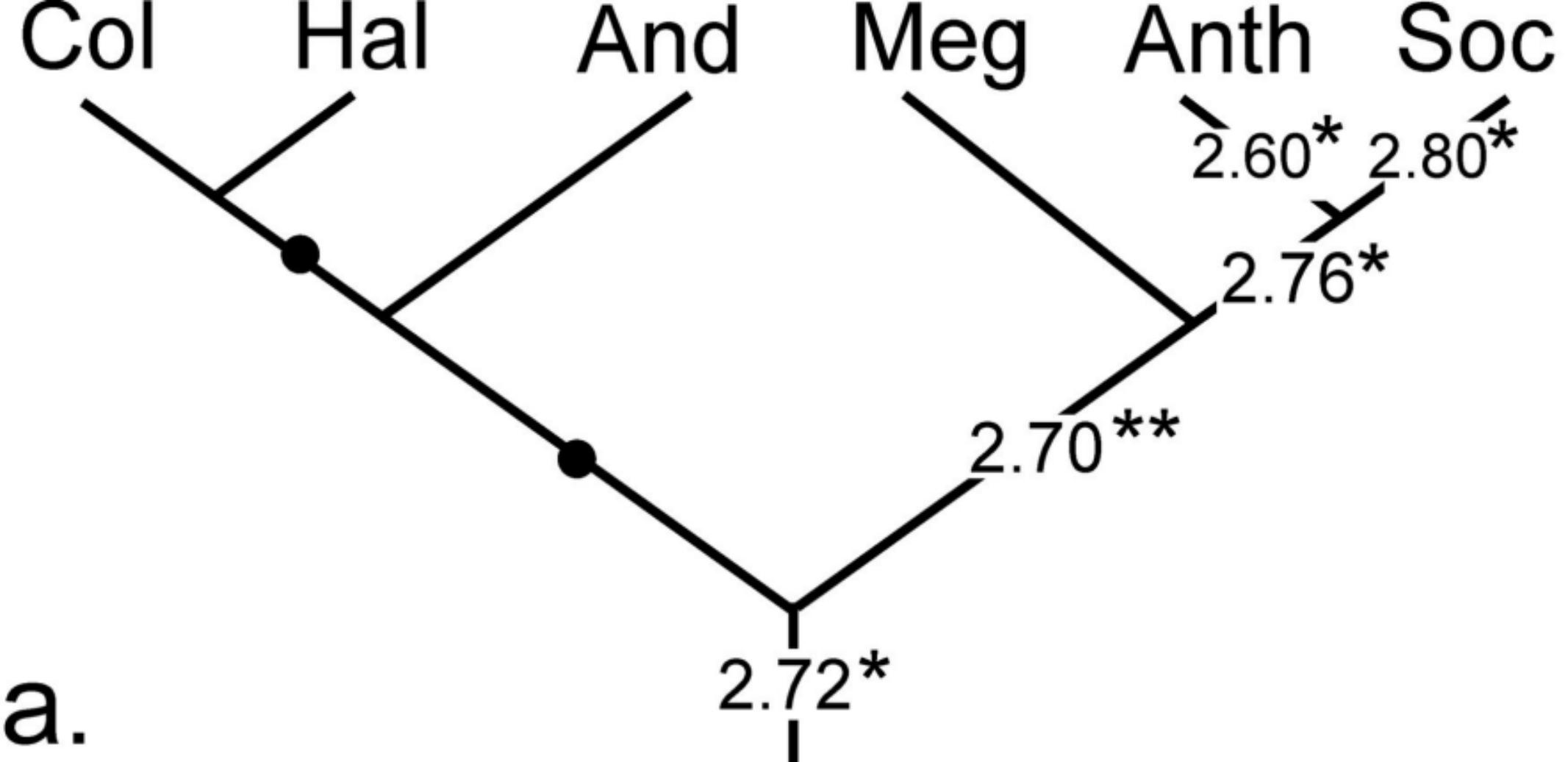

a.

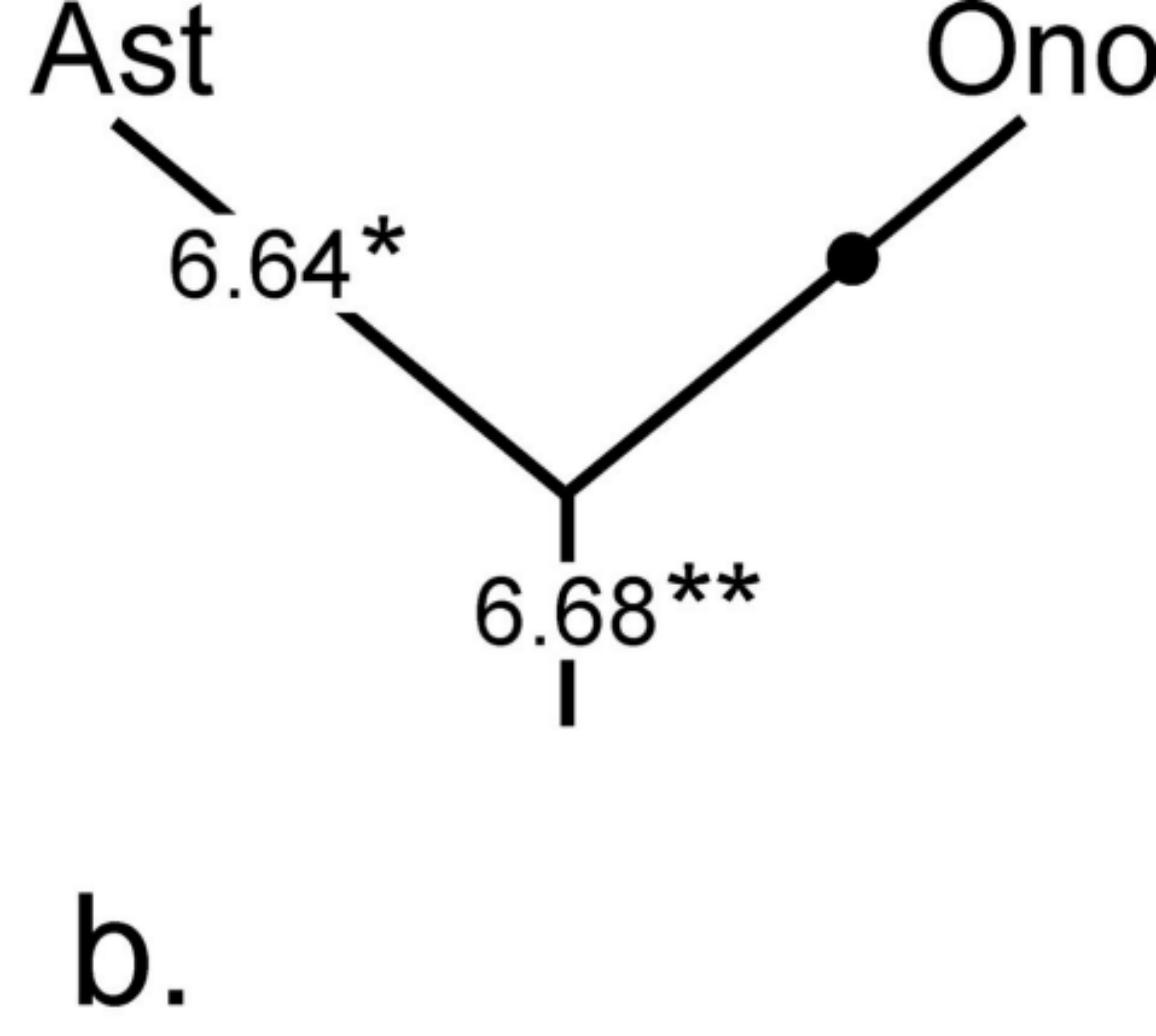

b.